\documentclass[12pt,preprint]{aastex}
\slugcomment{ApJ, in press}
\shortauthors{Geballe et al.}
\shorttitle{K-band Spectrum of IRS~8}
\begin{document} 

\title{The K-band Spectrum of The Hot Star in IRS~8: An Outsider in the
Galactic Center?\altaffilmark{1}}

\author{T. R. Geballe\altaffilmark{2}, F. Najarro\altaffilmark{3},
F. Rigaut\altaffilmark{2}, J.-R. Roy \altaffilmark{2}}

\altaffiltext{1}{Based on data obtained at the Gemini Observatory, which
is operated by the Association of Universities for Research in Astronomy,
Inc., under a cooperative agreement with the NSF on behalf of the Gemini
partnership: the National Science Foundation (United States), the Particle
Physics and Astronomy Research Council (United Kingdom), the National
Research Council (Canada), CONICYT (Chile), the Australian Research
Council (Australia), CNPq (Brazil) and CONICET (Argentina).}

\altaffiltext{2}{Gemini Observatory, 670 N. A'ohoku Place, Hilo, Hawaii
96720 USA; tgeballe@gemini.edu}

\altaffiltext{3}{Instituto de Estructura de la Materia, CSIC, Serrano 121,
29006 Madrid, Spain}

\begin{abstract}

Using adaptive optics at the Gemini North telescope we have obtained a
K-band spectrum of the star near the center of the luminous Galactic
center bowshock IRS~8, as well as a spectrum of the bowshock itself. The
stellar spectrum contains emission and absorption lines characteristic of
an O5-O6 giant or supergiant. The wind from such a star is fully capable
of producing the observed bowshock. However, both the early spectral type
and the apparently young age of the star, if it is single, mark it as
unique among hot stars within one parsec of the center.

\end{abstract} 

\keywords{shock waves -- stars: winds, outflows -- Galaxy: center --
infrared: ISM, stars}

\section{Introduction}

The nature of the Galactic center source IRS~8 \citep{bec75}, one of the
brightest compact mid-infrared sources in the central infrared cluster,
was unknown until adaptive optics H- and K-band imaging revealed that the
bulk of its infrared emission originates in a classic bowshock
\citep{rig03,geb04}. \citet{geb04} showed that the IRS~8 bowshock is a
straightforward consequence of the interaction of a dense and
high velocity wind from a hot star that is traversing moderately dense
interstellar gas. Adaptive optics imaging on large telescopes easily
resolves the central star of IRS~8 (hereafter IRS~8*) from the bowshock.

That stars undergoing mass loss within the interstellar medium in the
Galactic center produce bowshock-like structures of swept-up gas had
already been suggested from observations of IRS~21 by \citet{tan02} and,
by analogy, for a number of other luminous mid-infrared sources in the
Northern Arm. This has now been observationally verified \citep{tan05}
and additional examples of the phenomenon have been found
\citep{gen03,cle04a,cle04b}. Thus IRS~8 is the most graphic example of a
common phenomenon in the Galactic center that may also include a large
number of lower luminosity sources with less massive winds
\citep[e.g.,][]{eck04}.

Because of its large angular dimensions, fortuitous orientation, and
relatively isolated location, spectroscopy of IRS~8 at high angular
resolution can provide information both on the nature of the stellar
source of the wind and the properties of the wind and the interstellar
medium in the Galactic center.  Here we describe exploratory low
resolution K-band spectroscopy of IRS~8* and its bowshock.

\section{Observations and Data Reduction} 

K-band spectra of IRS~8 were obtained on UT 2004 June 9 and 2005 August
21 at the Frederick Gillett Gemini telescope on Mauna Kea, in
photometric conditions. The Gemini adaptive optics module ALTAIR was
used to feed NIRI, the facility near-infrared imager and spectrograph,
with a near-diffraction limited view of the IRS~8 region. The adaptive
optics correction was performed on a V$\sim$14~mag star 13~$\arcsec$
distant from IRS~8*. NIRI was configured with a 0.10~\arcsec\ slit and a
K grism to cover the 1.9-2.4~$\mu$m wavelength interval with resolving
power of 900. The plate scale was 0.0219\arcsec~pixel$^{-1}$. Exposures
were obtained with IRS~8* alternately positioned at two locations along
the slit, separated by 3$\arcsec$. The slit was oriented east-west on
the sky and thus intersected the bow shock a few tenths of an arcsecond
south-southeast of its apex, as shown in Fig.~1.

Spectra of both IRS~8* (covering 6 rows of the array) and the bow shock
(covering 7 rows of the array centered 0.24\arcsec\ east of the star were
extracted from the best quality subtracted pairs of flatfielded and
despiked spectral images, and were then coadded to produce final raw
spectra. Only the 2005 data consisting of 30 minutes of exposure time are
shown here. In the initial data reduction the spectra of IRS~8* and the
bowshock were ratioed by the spectrum of HR~7038 (F5V, observed
immediately after IRS~8), following artificial removal (via interpolation)
of the Br~$\gamma$ line in its extracted spectrum. Prior to ratioing a
slight smoothing was applied to all coadded spectra, resulting in a final
resolving power of 870, corresponding to a velocity resolution of
350~km~s$^{-1}$.  The wavelength scale was calibrated using telluric
absorption lines in the spectrum of HR~7038, and is accurate to
0.0003~$\mu$m (40~km~s$^{-1}$, 2$\sigma$).  

Examination of the intensity profile of the spectral images along the slit
indicated that contamination by the spectrum of the surrounding bowshock
contributed significantly to the spectrum of IRS~8*, e.g., 25$\pm$5
percent at 2.2~$\mu$m. This contamination was subtracted from the IRS~8*
spectrum to produce the final spectrum. As the spectrum of the bowshock is
smooth in the vicinity of almost all of the lines reported below in
IRS~8*, and because the bowshock is both fainter than IRS~8* at all but
the longest wavelengths in the spectrum and percentage-wise a minor
contaminant, the above uncertainty in the percentage of the bowshock
spectrum to be subtracted has a negligible effect on the strengths of
these lines and results in only a small increase in the noise level.

Figure 2 contains the reduced spectra of both the star and the bowshock.
Cancellation of the strong telluric band of CO$_{2}$ near 2.01~$\mu$m was
poor and thus that portion of the spectra is omitted in the figure. In
addition, as can be seen by the similar structures near the long
wavelength ends of the two spectra, accuracy in the 2.27-2.33~$\mu$m
regions of the spectra is limited by systematic errors in telluric line
removal. Over the bulk of each spectrum, however, the fluctuations appears
to be random.

The spectrum of the bowshock is a steeply rising continuum with weak and
narrow He~I singlet and H~I Br~$\gamma$ emission lines present near
2.059~$\mu$m and 2.166~$\mu$m, respectively. The spectrum of the central
star has a considerably less steep continuum and clearly contains several
weak emission lines in the 2.07-2.12~$\mu$m interval, but neither of the
emission lines seen in the bowshock. 

A more careful data reduction of the spectrum of IRS~8* was performed by
dividing the spectrum as before by the standard with the straight line
interpolation at Br~$\gamma$, but then multiplying by the normalized F5V
standard spectrum from \citet{wal97} at all wavelengths except in the
Br~$\gamma$ region prior to decontamination. By doing this subtle emission
features in the reduced spectrum resulting from weak absorption lines in
HR~7038 are removed. Because the strong Br~$\gamma$ lines in HR~7038 and
in the template do not quite match, we have used a straight line
interpolation there. The spectrum resulting from this reduction method is
used in the remainder of this paper.

From \citet{sco03} the extinction in the general vicinity of IRS~8 is
$A_V$$\sim$28~mag. When the spectrum of IRS~8* is dereddened by this
amount the continuum is noticeably redder than the Rayleigh-Jeans spectral
falloff of a hot star. Thus either the fully reduced and dereddened
stellar spectrum is still considerably contaminated by emission from dust
(despite our removal of contamination as described previously) or the
extinction to IRS~8* is somewhat larger than the value inferred from the
extinction map of \citet{sco03}. We suspect the latter and that fine
spatial structure in the extinction is the most likely cause of the
discrepancy. To estimate the extinction at IRS~8* we compared the
dereddened stellar spectral energy distribution (SED) with those from our
theoretical models of the IRS~8* spectrum (see below).  The best match was
obtained for a value of A$_{V}$=33~mag. Values differing from this by
$>$1.5~mag failed to reasonably reproduce the observed slope of the SED.
From our estimate of K=13.3 for IRS~8* based on these data\footnote{We
note the significant difference between the current K magnitude and the
one derived from photometry by \citet{geb04}, but use the present value
because of considerably improved performance of the Gemini adaptive optics
system since the previous measurement and because nothing is known about
the variability of IRS~8*.} we obtain a dereddened K=9.50~mag or
M$_k$=-5.0~mag for an assumed distance of 8~kpc to the Galactic center.
 
The reduced and dereddened 2.04-2.27~$\mu$m spectrum of IRS~8* is shown in
Fig.~3. The signal-to-noise ratio, estimated from the fluctuations in
lineless parts of the spectrum, varies with wavelength but is never far
from 100 in a resolution element ($\sim$0.0025~$\mu$m); it is highest from
2.21 to 2.27~$\mu$m. The narrow emission features near 2.1~$\mu$m have
peaks at 2.070~$\mu$m, 2.080~$\mu$m, 2.106~$\mu$m, and 2.116~$\mu$m.  
These features are respectively identified as: C IV (2.0705~$\mu$m; C IV
2.0796~$\mu$m); a blend of nearly coincident C III and N III lines at
2.1038~$\mu$m and 2.1081~$\mu$m; and nearly coincident C III, N III and O
III contributions at 2.1152~$\mu$m, 2.1155~$\mu$m, 2.1156~$\mu$m and
2.1162-69~$\mu$m \citep{naj95,han96}.  All wavelengths are {\it in vacuo}.
The He~I lines at 2.1126~$\mu$m and 2.1138~$\mu$m are very weak and
embedded within the emission feature at 2.116~$\mu$m (see discussion
below). An additional weak but statistically significant emission feature
is present at 2.247~$\mu$m and is identified as the stronger and shorter
wavelength member of an N~III doublet that has previously been seen in
late WN stars \citep{fig97,naj04} and perhaps is also weakly present in
the O stars observed by \citet{naj04}. The statistically significant
absorption feature at 2.190~$\mu$m is He II 10-7. The He I
2$^{1}$P-2$^{1}$S line is weakly in absorption at 2.059~$\mu$m.  All of
these features except for the last are also clearly detected in the
spectrum of IRS~8* obtained in 2004 June. The Br~$\gamma$ line is not
detected and its integrated equivalent width must be very small.

\section{Discussion}

\subsection{Radial velocities}

At the location of IRS~8 \citet{lac91} identified two velocity components
in the Ne~II line at 12.8~$\mu$m, one at -10~km~s$^{-1}$ which is
localized at IRS~8 and the other centered near +110~km~s$^{-1}$ associated
with the Northern Arm. The signal-to-noise ratios of the lines in the
IRS~8* spectrum are not high and the resolving power is low; hence the
radial velocities of the lines cannot be determined accurately.
Nevertheless, we can draw some limited conclusions from our model fits
(see section 3.3) to the observed spectrum. Fits to the 2.07-2.12~$\mu$m
complex of lines gives -10 ~km~s$^{-1}$. Fits to the He absorption lines
give +30 ~km~s$^{-1}$.  The mean value of these is +10~km~s$^{-1}$. The
strong Br~$\gamma$ line in the bowshock is centered at -40~km~s$^{-1}$.
All of these values match the radial velocity of the more negative
component of the Ne~II line to within the uncertainties. We thus conclude
that the blueshifted Ne~II component seen by \citet{lac91} is associated
with IRS~8* and suspect that it arises in gas that has been swept up by
the wind from IRS~8*.  Apparently, the Northern Arm is not interacting
with IRS~8 and thus IRS~8* lies either well in front of it or behind it.

\subsection{Classification of IRS~8*}

The relatively small equivalent widths of the emission lines in the
spectrum of IRS~8* indicate that the star is an OB type rather than a
Wolf-Rayet type. A very few WC9 stars with weak 2~$\mu$m lines have been
found by \citet{fig97}; however, the lines are considerably broader than
observed here. The isolated C~IV lines in IRS~8* have widths that are
only marginally broader than the velocity resolution of 350~km~s$^{-1}$,
indicating that the intrinsic full widths at half maximum (FWHMs) of the
lines are not larger than 200~km~s$^{-1}$, consistent with rotational
velocities found for O stars with lines originating at the photosphere.
The presence of the two helium lines in absorption suggests that the star
is either an Of or WNL type \citep{han96,fig97}, but the latter
classification is unrealistic as the carbon lines at 2.07~$\mu$m and
2.08~$\mu$m are prominent. Thus we are confident that we are observing an
O star.

\citet{han96}, and \citet{han05} have developed an infrared
classification scheme for OB stars, based on spectra in the H window and
short wavelength half of the K window, which is useful for hot stars that
suffer large extinctions.  Figure~3 is a comparison of the IRS~8* K-band
spectrum with online-available K-band spectra from the \citet{han96}
catalog for O stars ranging from O4 to O6.5 and different luminosity
classes. The resolving powers for all template spectra have been degraded
to 800 for direct comparison with the observed spectrum.

From Fig.~3 we judge that IRS~8* falls within the O5-O6.5 and III-If
ranges, with likely O5-O6~If spectral type and luminosity class.  For
stars earlier than O5-O6If the C~III/N~III lines become much weaker while
the strengths of the C~IV lines are considerably reduced for both earlier
and later spectral types. In addition, in cooler stars He I usually begins
to develop a noticeable absorption at 2.113~$\mu$m (see \citet{han96} for
further objects with later spectral type), but this line is not detected
in IRS~8*. A similar trend is observed in the He~I line at 2.059~$\mu$m
line, as its absorption strength clearly increases toward later types. Two
spectral features may be used to determine the luminosity class: the
emission feature at 2.116~$\mu$m and Br~$\gamma$. From Fig.~3 we see that
for a given spectral type the emission strength of the 2.116~$\mu$m
feature is larger for supergiants than for dwarfs. Also emission in the
strong stellar winds present in the supergiants starts to fill the
Br~$\gamma$ photospheric absorption profile to drive this line into
emission as opposed to the clean absorption profile observed for giants
and dwarfs.

Further constraints on the spectral type of IRS~8* might come from
measurements of the ionization state of the surrounding gas using
mid-infrared fine structure lines of ions with a diagnostic range of
ionization potentials. \citet{lac80} found from observations of the
Ne~II, Ar~III, and S~IV lines that the overall ionization state in the
central parsec of the Galaxy is consistent with excitation by stars with
$T_{eff}$~$\le$~35,000~K.  If that constraint applies to IRS~8 the
earlier O spectral subtypes in the above range could be ruled out. The
Ne~II line intensity at IRS~8 has been measured and is strong, but
searches for the Ar~III and S~IV lines at IRS~8 have not been reported.

\subsection{Stellar parameters and abundances}

Given the strong spectral similarities of IRS~8* with the O5-6 supergiants
in Cyg~OB2 (see Fig.~3), we computed model fits covering that parameter
domain, drawing from our analysis of the Cyg~OB2 stars for which UV,
optical and IR spectra are available \citep[][Najarro et al. in
preparation]{h02}. To perform quantitative analysis we utilized the
iterative, non-LTE line blanketing method presented by \citet{hillier98}
and proceeded as described in \citet{naj04}, also taking into account the
effects of the Fe~IV lines close to the wavelength of the He~I 2.06~$\mu$m
line \citep{naj06}. The reader is referred to \citet{hillier98} and
\citet{hillier99} for a detailed discussion of the code. Our best-fitting
model is displayed in Fig.~3 (dashed line) and reproduces the observed
K-band spectrum of IRS~8* quite well. Although the observed K band flux
and low resolution spectrum are insufficient to tightly constrain all of
the stellar properties of a supergiant such as IRS~8* (higher resolution
and other H lines would be required), it is possible to obtain accurate
estimates of some crucial parameters such as temperature and luminosity
and useful constraints on the wind density and metal abundances.

To determine the effective temperature we use He~I/II ionization balance
as the main constraint via the He~I 2.06 and 2.112/3~$\mu$m lines and the
He~II~2.189~$\mu$m line. The C~IV lines are considered as secondary
T$_{eff}$ indicators (see below).  The strengths of the absorption
components of neutral helium lines display the highest sensitivity to
changes in temperature, while the He~II 2.189~$\mu$m line also shows
strong sensitivity to wind density ($\dot M$).  Within the parameter
domain of interest, for reasonable He enrichment (e.g. He/H $<$ 0.25 by
number), these lines are not highly sensitive to changes in He abundance
if the mass loss rate is also adjusted to reproduce Br~$\gamma$. This may
be understood as follows. When the He abundance increases, the He
absorption lines become somewhat stronger.  However, to recover the
observed Br~$\gamma$ strength $\dot M$ must be increased to compensate for
the reduction in the hydrogen abunance. The higher mass loss rate refills
the He absorption components and, therefore, the resulting He profiles do
not change significantly.  This situation is reversed when the He
abundance is high enough (He/H~$>$~0.25 by number) as both effects,
enhancement of He abundance and increased $\dot M$, drive the He lines
into emission. From the above, we obtain T$_{eff}$=36000 $\pm$ 2000, 
log~$L$/L$_{\odot}$ = 5.6 $\pm$ 0.2 (set by the derived T$_{eff}$ and the
observed K magnitude), and 0.10~$<$~He/H~$<$0.25 by number.

The error estimate for the effective temperature may seem small, but we
are confident of the robustness of the derived value given the strong
sensitivity of the He~I absorption components and the He~II line together
with the presence of the C~IV lines. For temperatures above 38000~K the
He~I lines absorption components disappear while they clearly become too
strong for effective temperatures below 34000~K. Likewise, the He~II line
starts to fade for temperatures below our quoted range and gets too strong
in absorption for temperatures above it.  Is also important to note that,
for the relevant parameter domain, the He~I/II lines react only minorly to
changes on the surface gravity. The error in the stellar luminosity
reflects those in the effective temperature, extinction, contribution of
the bow-shock spectrum and the uncertainty in the flux calibration.
  
The Br~$\gamma$ line is strongly sensitive to wind density, clumping, and
velocity field, and to a lesser extent to gravity. Since we do not have
diagnostics for the terminal velocity to constrain the wind density, we
adopt a typical value, 2500~km~s$^{-1}$, found for other galactic O5If
stars \citep{h02}. Further, the spectral resolution is not high enough to
clearly constrain either the shape of the velocity field \citep[e.g., the
$\beta$ parameter][]{naj04} or the clumping factor ($f$). Thus, depending
on the adopted $\beta$ and He abundance we obtained clumping scaled mass
loss rates $\dot M / \sqrt{f}$ in the
range4.5--6.2~$\times~10^{-6}$~M$_\odot$~yr$^{-1}$. For our models we
assumed a clumping factor of $f$=0.15, based on our experience of modeling
the Cyg~OB2 objects, and hence the resulting mass-loss rates ranged from
1.75 to 2.45~$\times~10^{-6}~$M$_\odot$~yr$^{-1}$ for the adopted $\beta$
and terminal velocity. Likewise, the lack of cleaner diagnostic lines for
gravity such as the Brackett series lines in the H-band, hampers the
determination of log~g. Models with gravities above log~g=3.70 started to
fail to reproduce the spectrum. Reasonable fits were obtained for models
with log~g=3.40--3.60.

The C~IV lines were utilized as a secondary temperature diagnostic
\citep[see also][]{len04}, as they not only depend quite strongly on
temperature, but also on wind density, carbon abundance, gravity and the
$\beta$ parameter. If none of the above parameters are precisely
determined, the uncertainty in the carbon abundance may be as high as a
factor of three. In the case of IRS~8* currently acceptable values for the
carbon abundance range from 0.25$~\times$~solar to 0.6~$\times$~solar.

The N~III doublet at 2.25~$\mu$m reacts mainly to nitrogen abundance and
only slightly to effective temperature within the parameter domain of
interest and thus constitutes an excellent diagnostic of the nitrogen
abundance \citep[see also][]{naj04}.  However, the weakness of this
feature relative to the noise level results in a relatively high
uncertainty in the lower limit to the nitrogen abundance. We estimate an
enrichment of roughly 5~$\times$~solar with 7.5~$\times$~solar and
2.5~$\times$~solar as reasonable upper and lower limits respectively.

The strong emission feature at 2.116~$\mu$m in IRS~8* is of particular
interest. In the past this feature, which is present over a very wide
range of O spectral types and luminosities \citep{han96} has been
attributed to C~III and N~III n=8--7 transitions.  We tried to reproduce
this feature with the carbon and nitrogen abundances derived from the C~IV
2.07-2.08~$\mu$m and N~III 2.25~$\mu$m lines and missed more than half of
the observed equivalent width of the feature. Increasing either the carbon
or the nitrogen abundance to match the 2.116~$\mu$m feature resulted in
far too strong lines of C~IV at 2.07-2.08~$\mu$m and/or N~III at
2.25~$\mu$m. This result is confirmed by inspection of the dominant ions
of these species within the atmospheric region where the above lines form.
Carbon is largely divide between C~IV and C~V, so the C~III lines are too
weak even with a large increase in the carbon abundance. The dominant ions
of nitrogen and oxygen are N~IV and O~IV and hence the N~III and O~III
recombination lines are reliable diagnostics for deriving
abundances. Hence, we now believe that the 2.116~$\mu$m feature in IRS~8*
is dominated by O~III n=8--7 transitions. After extending our O~III
model atom to account for these transitions we could satisfactorily
reproduce the observations (see Fig.~3). Further, the O~III component of
the 2.116~$\mu$m feature largely depends on the oxygen abundance and only
slightly on gravity, effective temperature, wind density, and velocity
field. Thus, this feature may be a powerful diagnostic of oxygen
abundance, and therefore an important metal abundance determiner, over a
wide range of O spectral types (Najarro et al. in preparation).  Using it
we obtain an oxygen abundance of 0.8 to 1.1$~\times$~solar in IRS~8*,
which indicates solar metallicity for the cloud in which IRS~8* formed.

To summarize this subsection, the stellar parameters T$_{eff}$, L, $\dot M
/ \sqrt{f}$, and log g are fully compatible with OIf supergiants
\citep[e.g.][]{h02} and confirm the nature of IRS~8* derived from
classification schemes.  The derived N/C and N/O ratios reveal partial CNO
processing at the surface of IRS~8 and are consistent with the values
expected for an O supergiant if rotational mixing has taken place.

\citet{geb04} used the measured standoff distance of the bowshock from the
star together with a rough estimate of the particle density
($n$~=~10$^{3}$~cm$^{-3}$) of the ambient interstellar medium in the
Galactic center and typical values of windspeed (10$^{3}$~km~s$^{-1}$ and
mass loss rate in a hot star (10$^{-6}$M$_\odot$~yr$^{-1}$) to estimate a
space velocity $v_*$ for IRS~8* of 150~km~s$^{-1}$, which is a reasonable
value in the Galactic center. The standoff distance is proportional to
($\dot{M}$~$v_w$~/~$n$~~$v_*$$^{2}$)$^{0.5}$, where $\dot{M}$ is the mass
loss rate and $v_w$ is the wind speed. The values of $v_w$ and $\dot{M}$
used in the above modelling suggest that in order to produce the observed
standoff distance $n$ would need to approach 10$^{4}$~cm$^{-3}$, which
would not be surprising.

\subsection{Evolutionary status and mass of IRS~8*: An outsider within the
Galactic center context?}

Recently, \citet{paum06} have reported the spectroscopic identification
of $\sim$40 OB supergiants, giants and dwarfs in the central parsec of
the galaxy. Interestingly they find no OB stars outside the inner 0.5~pc
(radius) of the galaxy and the earliest spectral type in their OB sample
lies around O8-9I. They derive a common age of $6~\pm~2~$Myr for the
cluster. 

Our analysis suggests that IRS~8*, although only 1~pc from the center,
does not fit into this picture of the central cluster of hot stars. It is
of much earlier spectral type than any of the stars classified by
\citet{paum06}. Currently it is the only known OB star outside the
central 0.5~pc region of the cluster. Figure~4 shows the position of
IRS~8* (solid cross) as estimated from our model fits in the HR diagram
compared with different evolutionary scenarios. An estimate of the
evolutionary status of IRS~8*, based on comparisons with evolutionary
tracks of stars without rotation \citep[e.g.][not displayed in
Fig.~4]{madx} yields a star with a zero age main sequence (ZAMS) mass of
48~M$_\odot$ and a current age of 2.8~Myr.  Such a star would not show any
processed material on its surface.  This is clearly at odds both with the
current estimate for the age of the Galactic center cluster and with the
abundance pattern derived from our models. The situation improves when
evolutionary models accounting for rotation (dashed-lines in Fig.~4) are
considered \citep{madx}. Then the current position in the HR diagram
corresponds to a star with a ZAMS mass of 44.5~M$_\odot$ and an age of
3.5~Myr. Such a star would show CNO-processed material on its surface.
Except for the age, still well below the estimate obtained by
\citet{paum06} using non-rotating models for a single burst scenario, the
stellar parameters, including the abundance pattern, are fully consistent
with those derived from our modelling.

The crucial question thus is whether this star is really much younger than
the cluster and probes the existence of ongoing (or at least much more
recent) star formation, or if on the contrary the star is either an
impostor or a cluster member that underwent a rejuvenation cure. A
possible way out is provided if the star originally was a member of a
massive close binary system. In such a case, we could be looking now at
the secondary star, with the primary either exploded as supernova or in an
evolutionary phase when it is much dimmer at $K$ than the secondary.
Models for massive close binaries have been developed by \citet{well99} to
explain the optical counterparts of massive X-ray binaries. Using these
models we have found that for a massive close binary system with initial
masses of 25~M$_\odot$ and 24~M$_\odot$ (their model 10a) the current
position of IRS~8* may be elegantly explained without violating the age of
the Galactic center cluster. Similar scenarios are a possible explanation
for some of the overluminous He~I objects in the central parsec.  The
solid lines in Fig.~4 correspond to the evolution of the primary (grey)
and secondary (black) components of the massive close binary system. The
thick solid line displayed on the track of the secondary corresponds to
the phase where the primary is at least 10 times less bright in $K$ than
the secondary. The observed location of IRS~8* in the HR diagram is
reached after 7.1~Myrs, which is consistent with the $6~\pm~2$~Myr
estimate from \citet{paum06}. Furthermore, at this stage the surface
enrichment displayed by the secondary shows excellent agreement with the
values derived in our model. For this particular massive close binary
model the primary has not exploded yet, as otherwise the secondary would
be already halfway through core helium burning and would have a much lower
effective temperature. However, models with a slightly more massive
primary and a slightly less massive secondary (N. Langer, private
communication) could also reproduce the elemental abundances and current
position of IRS~8* in the HR diagram after the explosion of the primary
(at $\sim$6~Myr) and, in addition, could have provided a kick to the
secondary, placing it at its current position outside the central 0.5~pc.

The spectroscopic mass of IRS~8* is in the range 23--37~M$_\odot$ (for
log~g 3.40--3.60), although the range could be larger, bearing in mind the
uncertainties in determining the stellar gravity, as discussed previously.  
From the evolutionary models of \citep{madx}, the current mass of IRS~8*
is 38~M$_\odot$ for a single star evolution with rotation, and thus
consistent with the highest values of spectroscopic mass. Likewise, in the
close binary scenario, the current mass of IRS~8* (after consuming a
significant fraction of its companion) is 36--40~M$_\odot$.

Future spectra at higher resolution than presented here should be able to
test if IRS~8* is part of a close binary. If IRS~8* is single its
origin is highly uncertain. It is then either an impostor in the central
parsec or it is a loosely associated member of the central cluster of hot
and massive stars. The current motion of IRS~8*, nearly directly away from
the center (as judged by the orientation of the bowshock and low radial
velocity), suggests that it passes close to Sgr~A* \citep{geb04} and may
once have been a member of central cluster. If so it might be that a small
population of less evolved and less luminous mid-late O type stars with
weak emission lines such as those of IRS~8* are still hidden within the
cluster, which could challenge the current understanding of that cluster
as having a common age \citep{paum06}.

\begin{acknowledgements}

We thank the staff of the Gemini Observatory for its support of these
observations, which were performed as part of program GN-2004A-SV-203. We
also thank the builders of the ALTAIR adaptive optics system at Gemini.
F.N. acknowledges PNAYA2003-02785-E AYA2004-08271-C02-02 projects and the
Ramon y Cajal program. We thank John Hillier for providing his atmospheric
code. We also are indebted to Keith Butler for his advice on upgrading the
model atoms and especially to Norbert Langer for fruitful discussions and
providing the evolutionary tracks for massive close binaries. We are
grateful for advice from Paul Crowther and Don Figer and for helpful
comments from the referee.

\end{acknowledgements}

\clearpage
\begin{figure}
\epsscale{0.9}
\plotone{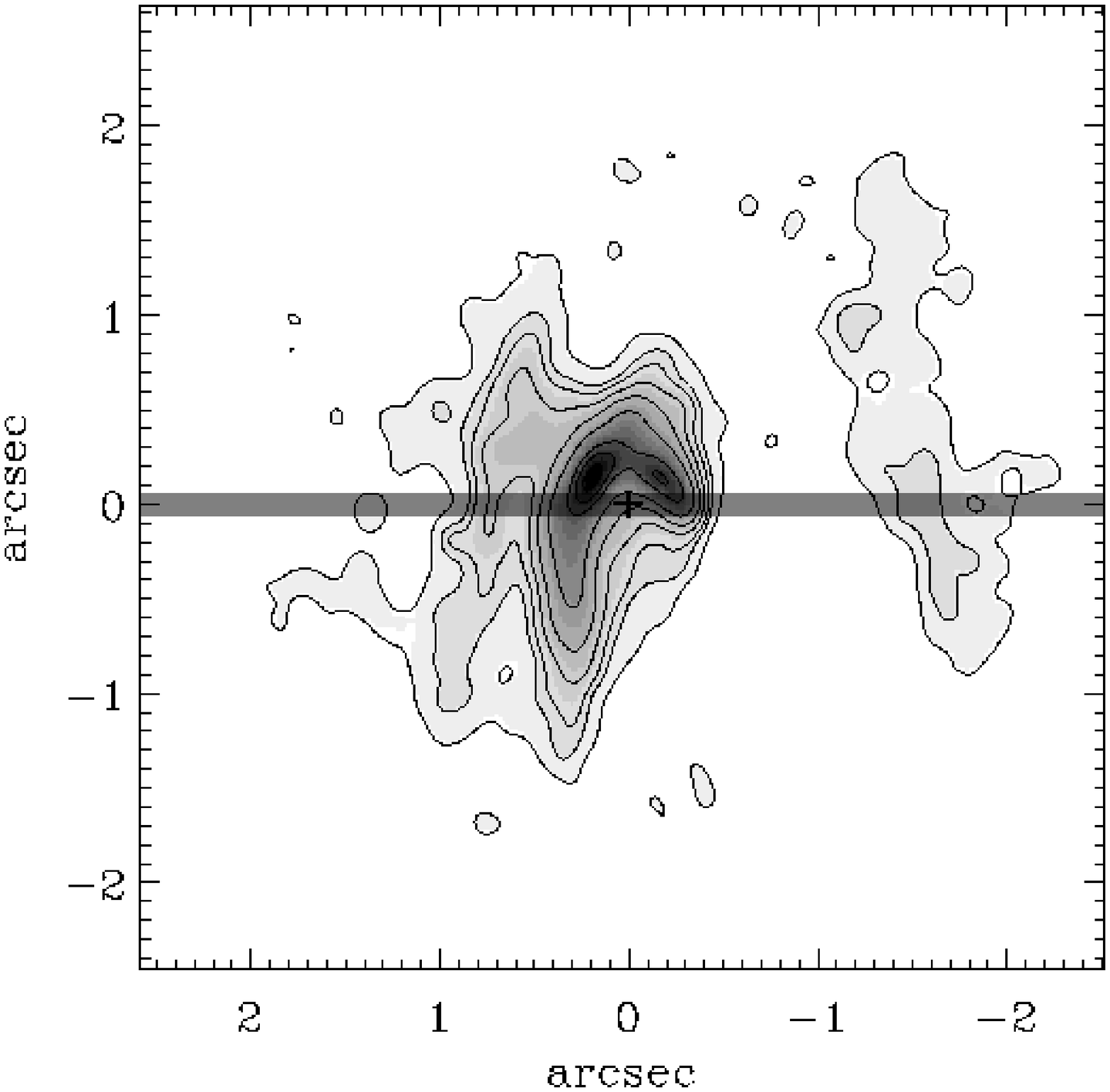} 
\caption{Contour plot of the IRS~8 region obtained through a narrow
band 2.3~$\mu$m filter, with the point sources removed, from
\citet{geb04}.  The cross indicates the location of the central star of
IRS~8. North is up; east is to the left. The location of the slit used to
obtain the spectra presented here is denoted by the narrow shaded
rectangle.}
\label{fig1}
\end{figure}

\clearpage

\begin{figure} \epsscale{1} \plotone{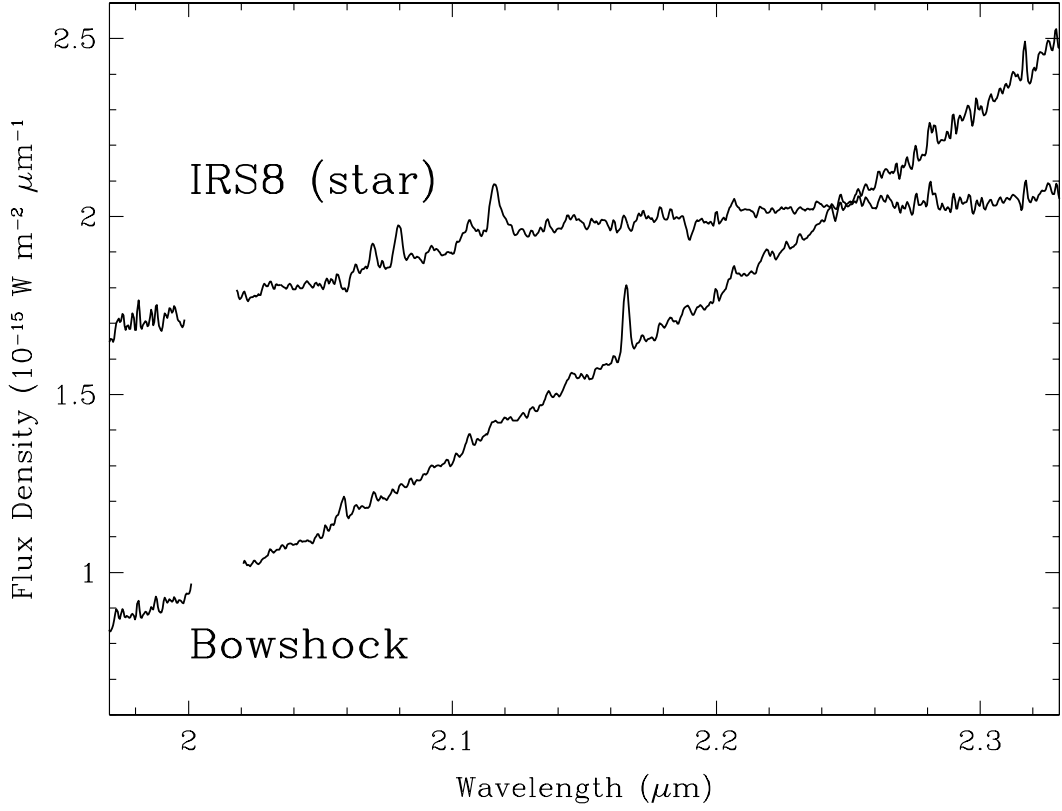} \caption{Spectra of the
central star of IRS~8 and of a 0.10''~$\times$~0.15''
(NS$\times$EW) portion of the bowshock centered 0.24'' east of the central
star.}
\label{fig2}
\end{figure}

\clearpage

\begin{figure}
\includegraphics[scale=0.65,angle=90]{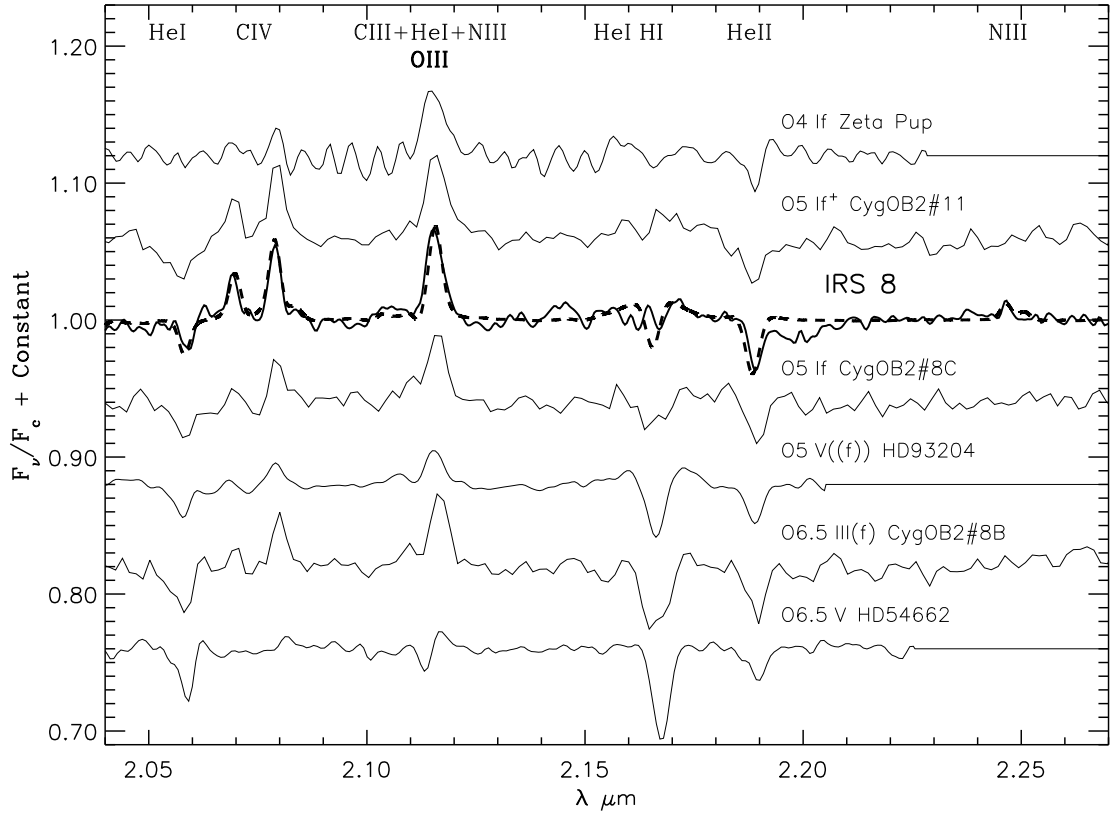}
\caption{ Spectral type determination of IRS~8*. Comparison of the 
resulting normalized spectrum with K-band spectra from \citet{han96}.
degraded to a resolution of R=800. Also displayed (dashed) is a model fit
with stellar parameters corresponding to an O5.5If star (see
text). Wavelengths of important emission and absorption lines are
roughly indicated by labels at top of figure} \label{fig3}
\end{figure}

\begin{figure} \includegraphics[scale=0.65,angle=90]{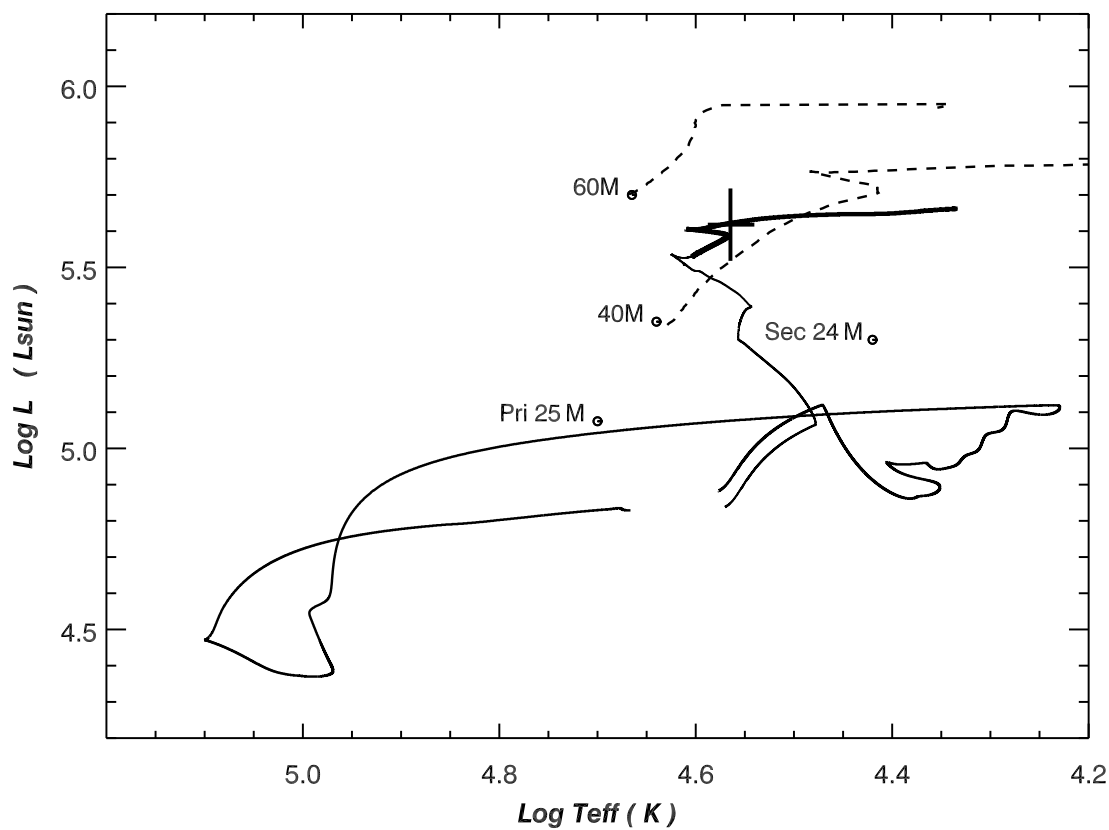}

\caption{Position of IRS~8* (solid cross) in the HR diagram as estimated
from model fits, compared with different evolutionary scenarios. Dashed
curves correspond to models for 60 and 40M$_\odot$ single stars including
rotation \citet{madx} during the pre-WN phase. Solid curves correspond to
the evolution of the primary and secondary components of a
massive close binary system with initial masses of 25 and 24M$_\odot$
\citet{well99}, their model 10a.  The thick solid line displayed on the
track of the secondary corresponds to the phase where the primary is at
least 10 times less bright at K than the secondary. The current location
of IRS~8* is reached after 3.6 and 7.1 million years for the single star
and massive close binary evolutionary cases respectively (see discussion
in text).} 
\label{fig4} 
\end{figure}

\end{document}